\documentclass[journal]{IEEEtran}
\usepackage{amsmath}
\usepackage{amsthm,amssymb}
\usepackage{graphicx}
\usepackage{subfigure}
\usepackage{booktabs}
\usepackage{comment,epsfig}
\usepackage[linesnumbered,ruled,vlined]{algorithm2e}

\ifCLASSINFOpdf

\else

\fi

\newtheorem{proposition}{Proposition}

\begin{document}

\title{A Low Complexity Encoding Algorithm for Systematic Polar Codes}

\author{Guo Tai Chen,~
        Zhaoyang Zhang,~
        Caijun Zhong,~%
        and~Liang Zhang
\thanks{G. T. Chen (email: chenguot@163.com) is with Fuqing Branch of Fujian Normal University, China, and he is also a domestic visitor at Zhejiang University from 2015 to 2016. Z. Zhang (Corresponding Author, email: ning\_ming@zju.edu.cn), C. Zhong (email: caijunzhong@zju.edu.cn) and L. Zhang are with the College of Information Science and Electronic Engineering, Zhejiang University, China.}\\
\thanks{This work was supported in part by the National Key Basic Research Program of China under Grant 2012CB316104, the National Natural Science Foundation of China under Grants 61371094 and 61401099, the Huawei HIRP Flagship Projects under Grants YB2015040053 and YB2013120029, Zhejiang Provincial Natural Science Foundation of China under Grant LR15F010001, the Fundamental Research Funds for the Central Universities under Grant 2016QNA5004, the Research Funds of the Education Department of Fujian Province under Grants JA12350 and JA14339.}
}

\markboth{Submitted to IEEE Communications Letters, February~2016}%
{Shell \MakeLowercase{\textit{et al.}}: Bare Demo of IEEEtran.cls for IEEE Journals}

\maketitle

\begin{abstract}
Ar\i kan has shown that systematic polar codes (SPC) outperform nonsystematic polar codes (NSPC). However, the performance gain comes at the price of elevated encoding complexity, i.e., compared to NSPC, the available encoding methods for SPC require higher memory and computation. In this letter, we propose an efficient encoding algorithm requiring only $N$ bits of memory and having $\frac{N}{2}\log_2N$ XOR operations. Moreover, the auxiliary variables in the algorithm can share the memory to reduce extra memory requirement. Furthermore, a parallel 2-bit encoding algorithm is also presented to improve the encoding throughput. Remarkably, we show that parallel encoding can be implemented with the same number of XOR operations and memory bits. Finally, the proposed encoding algorithm can be directly used for NSPC with the same complexity.
\end{abstract}


\IEEEpeerreviewmaketitle

\section{Introduction}
Polar codes, originally proposed by Ar\i kan in \cite{Arikan2009}, have gained enormous interests due to a number of distinctive features. For instance, polar codes have explicit coding structure and can achieve the capacity of symmetric binary memoryless channels (S-BMC). Moreover, polar codes with finite length yield competitive performance when compared to LDPC \cite{Tal2015} and Turbo codes \cite{Niu2013} in addition to having low encoding and decoding complexity.

The standard polar codes are in nonsystematic form where both frozen bits and information bits (also referred to as user bits) are placed on the polarized bit-channels of the polarization structure and the user bits do not appear in the polar codeword. However, information bits as part of the codeword are required in some scenarios, such as the famous Turbo codes \cite{TC1996} whose component codes are systematic codes that can exchange information between modules in turbo decoding. To construct systematic polar codes (SPC), Ar{\i}kan proposed the idea of shifting the user bits from polarized bit-channels to unpolarized bit-channels \cite{Arikan2011}, which makes the frozen and user bits lie on two different extremes of polarization structure. Ar{\i}kan showed that systematic polar codes outperform nonsystematic polar codes (NSPC) in terms of bit error ratio (BER) and the performance have also been investigated in \cite{Liping2015}.

Recently, SPC as component codes of concatenated codes have been investigated in \cite{Aijun2016EL} and \cite{Aijun2016TCOM}. Compared to the NSPC with the same polarization structure, SPC is inherently more complex. Hence, to facilitate the application of SPC, the key challenge is to find an efficient encoding method. In \cite{Arikan2011}, Ar\i kan presented a recursive method for SPC encoding with $\alpha \frac{N}{2}\log_2N$ $(\alpha>1)$ XOR operations, where Ar\i kan also suggested using successive cancellation (SC) decoder as an encoder for SPC. Following this suggestion, an SPC encoder which facilitates easy parallelization was  proposed in \cite{Sarkis2014,Sarkis2015}, with the limitation of executing SC algorithm twice and constrained frozen bits. Another SPC encoding algorithm in the recursive implementation with elimination method was presented in \cite{Presman2012}. Most recently, the authors of \cite{Vangala2015} proposed three encoding algorithms for SPC with memories of $N(1+\log_2N)$, $2N-1$ and $N$ bits and XOR operations of $\frac{N}{2}\log_2N$, $N(1+\log_2N)$ and $N(1+2\log_2N)$, respectively. However, the major drawback of the above discussed encoding methods is the high requirements on memory or computation, which may not be suitable for devices with small size and limited power.
\begin{table}[tp]
\scriptsize
\caption{Summary of Systematic Polar Encoders}
\label{ComplexityTable}\centering %
\begin{tabular}{|c|c|c|c|}
\hline
Algorithm & Recursion & $\#$ bits(excl. I/O) & $\#$ XORs \\ \hline
EncoderA \cite{Vangala2015} & No & $N(1+\log_2N)$ & $\frac{N}{2}\log_2N$ \\ \hline
EncoderB \cite{Vangala2015} & Yes & $2N-1$ & $N(1+\log_2N)$ \\ \hline
EncoderC \cite{Vangala2015} & Yes & $N$ & $N(1+2\log_2N)$ \\ \hline
NSPC \cite{Vangala2015} & Yes/No & $2N$ & $\frac{N}{2}\log_2N$ \\ \hline
Proposed SPC & No & $N$ & $\frac{N}{2}\log_2N$ \\ \hline
\end{tabular}
\normalsize
\end{table}
Motivated by this, in this letter, we propose a new efficient encoding algorithm for SPC requiring only $N$ bits of memory (excluding the input/output) and $\frac{N}{2}\log_2N$ XOR operations. To the best of the authors' knowledge, the proposed algorithm requires the minimum memory as well as XOR operations compared to the known encoding methods, as illustrated in Table \ref{ComplexityTable}. In addition, to further improve the encoding throughput, a parallel 2-bit encoding algorithm is also discussed, which shows that the parallel encoding can be accomplished without incurring additional cost in terms of XOR operation and memory bit.
\section{Nonsystematic and systematic polar codes}
\begin{figure}[h]
\centering
\includegraphics[width=82mm]{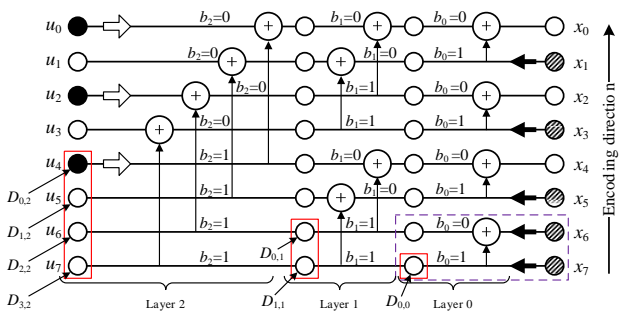}
\caption{SPC encoding diagram with $N=2^3$ and $\mathcal{A}=\{1,3,5,6,7\}$.}
\label{EncodingStructure}
\end{figure}

For polar codes with codeword length $N(=2^n,~n\geq 1)$ and kernel matrix $F=\begin{pmatrix}1&0\\1&1\end{pmatrix}$, the polarization transformation matrix $G$ can be written as $G=F^{\otimes n}$ where $\otimes$ denotes the Kronecker power operation. Let $\textbf{u}=(u_0,u_1,\cdots,u_{N-1})$ and $\textbf{x}=(x_0,x_1,\cdots,x_{N-1})$ be the bit vectors on the left and right side of the encoder shown in Fig.\ref{EncodingStructure}, respectively, then we have
\begin{equation}
\textbf{x}=\textbf{u}G.
\label{eqNSPC}
\end{equation}

For NSPC, $\textbf{u}$ is the only input of the encoder, i.e., $\textbf{u}$ includes both the frozen and user bits, and the encoding is performed from left to right according to the coding structure shown in Fig.\ref{EncodingStructure}, where $\mathcal{A}$ denotes the index set of the user bits.

However, for SPC, both $\textbf{u}$ and $\textbf{x}$ are inputs of the encoder. To construct systematic polar codes, Ar\i kan proposed to place the user bits on the right side of the encoder and keep the same indices for the bits as illustrated in Fig.\ref{EncodingStructure} with $N=2^3$ and $\mathcal{A}=\{1,3,5,6,7\}$, where the left extreme node with hollow arrow denotes the frozen bit while the right extreme node with solid arrow denotes the user bit.

The index set for the frozen bits is the complementary set of $\mathcal{A}$, i.e., $\mathcal{A}^c=\{0,1,\cdots,N-1\}-\mathcal{A}$. Now, denote $\textbf{u}_\mathcal{A}$ and $\textbf{u}_{\mathcal{A}^c}$ as the bit vector with elements $u_i$, $i\in \mathcal{A}$ and $i\in \mathcal{A}^c$, respectively, and the similar denotation is also for $\textbf{x}_\mathcal{A}$ and $\textbf{x}_{\mathcal{A}^c}$, then, Equation (\ref{eqNSPC}) can be rewritten as
\begin{equation}
(\textbf{x}_\mathcal{A}\ \textbf{x}_{\mathcal{A}^c})=(\textbf{u}_\mathcal{A}\ \textbf{u}_{\mathcal{A}^c})\begin{pmatrix}G_{\mathcal{A}\mathcal{A}}&G_{\mathcal{A}\mathcal{A}^c}\\ G_{\mathcal{A}^c\mathcal{A}}&G_{\mathcal{A}^c\mathcal{A}^c}\end{pmatrix},
\label{eqSPC}
\end{equation}
where $G_{\mathcal{A}\mathcal{A}^c}$ is a sub-matrix of $G$ with elements $G_{i,j}$, $i\in \mathcal{A}$ and $j\in \mathcal{A}^c$, and $G_{\mathcal{A}\mathcal{A}}$, $G_{\mathcal{A}^c\mathcal{A}}$ and $G_{\mathcal{A}^c\mathcal{A}^c}$ are defined in the same fashion. The objective of SPC encoding is to obtain $\textbf{x}_{\mathcal{A}^c}$ given the inputs $\textbf{u}_{\mathcal{A}^c}$ and $\textbf{x}_{\mathcal{A}}$.

Since matrix $G_{\mathcal{A}\mathcal{A}}$ is invertible, $\textbf{x}_{\mathcal{A}^c}$ can be computed by \cite{Arikan2011}:
\begin{equation}
\textbf{x}_{\mathcal{A}^c}=(\textbf{x}_{\mathcal{A}}+\textbf{u}_{\mathcal{A}^c}G_{\mathcal{A}^c\mathcal{A}})G_{\mathcal{A}\mathcal{A}}^{-1}G_{\mathcal{A}\mathcal{A}^c}+\textbf{u}_{\mathcal{A}^c}G_{\mathcal{A}^c\mathcal{A}^c}.
\label{eqXf}
\end{equation}

As discussed in the Introduction section, the known methods in literature to compute $\textbf{x}_{\mathcal{A}^c}$ requires relatively large memory and heavy computation. Motivated by this, the main objective of this letter is to find an efficient algorithm to compute $\textbf{x}_{\mathcal{A}^c}$.
\section{Efficient encoding algorithm for SPC}
The proposed encoding algorithm is similar to the algorithm \textbf{EncoderA} in \cite{Vangala2015} where the encoding is implemented from the bottom horizontal connection to the top horizontal connection and the calculation for each horizontal connection starts from known node (i.e., one of elements of $\textbf{x}_\mathcal{A}$ or $\textbf{u}_{\mathcal{A}^c}$) and moves from one node to the next and to the other side of the polarization structure, as shown in Fig.\ref{EncodingStructure}. Due to the fact that each node is allocated with one bit memory in \textbf{EncoderA}, the total memory requirement of \textbf{EncoderA} is $N(1+\log_2N)$ bits. The key feature of the proposed encoding algorithm is to reduce the memory requirement from $N(1+\log_2N)$ bits to $N$ bits while maintaining the same computation load, i.e., $\frac{N}{2}\log_2N$ XOR operations.
\subsection{Encoding algorithm}
To exploit the recursive nature of polar codes, the polarization structure with $N=2^n$ is divided into $n$ layers labeled by $0,1,\cdots,(n-1)$ from right to left as shown in Fig.\ref{EncodingStructure}. And for layer $\lambda~(\lambda=0,1,\cdots,n-1)$, the $N$ nodes (includes the corresponding operations) are separated into $2^{(n-1)-\lambda}$ blocks from top to bottom and each block contains $2^{\lambda+1}$ elements. As an illustration, the dashed box in the right-bottom corner of Fig.\ref{EncodingStructure} represents one of the first-layer blocks.

To analyze the memory requirement of the encoding algorithm, let us first consider the blocks in the same layer. The key observation is that the blocks in same layer are independent and there is no information exchange between the blocks in the same layer, which indicates that the memory used for one block can be recycled for the other blocks in the same layer when the encoding proceeds from the bottom up. In addition, a close inspection reveals that, in each block, only the lower half of the elements need to be stored in memory for the encoding process, and the outcomes of the XOR operations in the upper half can be stored in the associated lower half. Then, it is easy to show that only $2^{\lambda}$ bits of memory are required for the encoding process in layer $\lambda$. Hence, the total required memory of all layers is $2^n-1=N-1$ bits.

To corroborate the above argument, let us consider the following illustrative example. For notational convenience, we define $D_{a_{\lambda},\lambda}$ as the  memory address used for layer $\lambda$, where $a_{\lambda}~(a_{\lambda}=0,1,\cdots,2^{\lambda}-1)$ is the index of bit memory. We focus on the bottom block in layer $0$ highlighted by the dashed box in Fig. \ref{EncodingStructure}. In this block, $x_6$ and $x_7$ are the known bits and $x_7$ will be first processed. The first step is to copy $x_7$ to $D_{0,0}$, and then to $D_{1,1}$ and $D_{3,2}$. Since the stored value in $D_{0,0}$ is the same as $x_7$, the XOR operation between $x_6$ and $x_7$ can be replaced with $x_6$ and $D_{0,0}$ and the outcome of the XOR operation can be stored in $D_{0,0}$, since the previous value in $D_{0,0}$ is obsolete. The next step is to copy the value of $D_{0,0}$ to $D_{0,1}$. Once done, $D_{0,0}$ is released since its value is no longer required in the remaining process, hence can be recycled for use in the next block in layer 0. As such, only $1=(2^0)$ bit memory is required for the encoding process in layer 0. Finally, the above process is extended to other layers.

In the encoding process of SPC, there exist two different operations, namely, directly copying and XOR operation. Hence, it is of significant interest to obtain a fast method to determine the proper operation. Here, we present a simple method to address this issue. Let $\phi~(\phi=0,1,\cdots,N-1)$ denote the index of current horizontal connection for top to bottom and denote the binary expression of $\phi$ as $b_{n-1}\cdots b_0$. Then, directly copying is performed in layer $\lambda$ when $b_{\lambda}=1$ and XOR operation occurs when $b_{\lambda}=0$, as illustrated in Fig. \ref{EncodingStructure}.

For the propagation from left to right, the case for $b_{\lambda}=0$ is more complex. Let us take the information propagation of $u_0$ for example. Since both $u_0$ and $D_{0,2}$ are needed for the XOR operation $u_0\bigoplus D_{0,2}$ at the same time, we can not copy $u_0$ to $D_{0,2}$. To circumvent this problem, we introduce a temporary variable $t$, and set $t=u_0$. Hence, the XOR operation $x_0\bigoplus D_{0,2}$ can be replaced with $t\bigoplus D_{0,2}$, and the corresponding outcome can be assigned to $t$ as well, i.e., $t=t\bigoplus D_{0,2}$. When $b_{\lambda}=1$, the directly copying operation is to copy $t$ into $D_{a_\lambda,\lambda}$, which implies that $t$ and $D_{a_\lambda,\lambda}$ share the same value after the directly copying, hence $t$ can then be used for the following operations instead of $D_{a_\lambda,\lambda}$. Due to the introduction of a temporary variable, the total required memory bits for the proposed encoding process is $N$. It is also worth emphasizing that the number of XOR operations in the proposed encoding process is only $\frac{N}{2}\log_2N$.

The pseudocodes of the proposed encoding of SPC is listed in \textbf{Algorithm 1} where $\leftarrow$ is the assignment operator. Lines 6-15 are for the propagation from right to left, and lines 17-27 are for the propagation from left to right. After each process of propagation, $a_{\lambda}$ will be updated from the next propagation, which is shown in lines 28-31.

Comparing the pseudocodes between \textbf{EncoderA} in \cite{Vangala2015} and \textbf{Algorithm 1}, it can be found that the encoding processes of both algorithms work in the same serial fashion and are implemented from horizontal connection $(N-1)$ to horizontal connection 0 one by one, which indicates that both algorithms have the same number of XOR operations and directly copying. However, our proposed algorithm repeatedly utilizes the $N$-bit memory while \textbf{EncoderA} requires $N(1+\log_2N)$ bits of memory. Moreover, the XOR operation in the proposed algorithm only requires two operands and is performed in place while the XOR operation in \textbf{EncoderA} has three operands including one destination and two sources, which may incur extra computation. And we will show in the next subsection that the updating of $b_{\lambda}$ and $a_{\lambda}$ does not need extra computation. Therefore, the efficiency of the proposed encoding algorithm has not degraded in comparison to \textbf{EncoderA} in \cite{Vangala2015}.

It is also worth pointing out that the proposed algorithm can also be used as an encoding algorithm for NSPC where the encoding only has the propagation from left to right based on the polarization structure similar to Fig.\ref{EncodingStructure}. This indicates that the minimum requirement of NSPC encoding is also $ \frac{N}{2}\log_2N$ XOR operations and $N$ bits of memory.
\begin{algorithm}[ht]
\small
\caption{Proposed Encoding algorithm for SPC}
\KwIn{$\textbf{u}$ and $\textbf{x}$ with unfilled bits (variables of (1));}
\KwOut{Full codeword $\textbf{u}$ and $\textbf{x}$;}
\label{alg:MyEncodingSPC}
\For(~~~//initialization){$\lambda=0:(n-1)$}{
    $a_{\lambda}\leftarrow 2^{\lambda}-1$\;
}
\For{$\phi=(N-1):-1:0$}{
    Store binary expression of $\phi$ into $b_{n-1}\cdots b_0$\;
    \If(~~~~~~~~~//information bits){$\phi\in\mathcal{A}$}{
        \If{$b_0=0$}{
            $D_{0,0}\leftarrow D_{0,0} \oplus x_{\phi}$\;
        }
        \Else{
            $D_{0,0}\leftarrow x_{\phi}$\;
        }
        \For{$\lambda=1:(n-1)$}{
            \If {$b_{\lambda}=0$}{
                $D_{a_{\lambda},\lambda}\leftarrow D_{a_{\lambda},\lambda}\oplus D_{a_{\lambda-1},\lambda-1}$\;
            }
            \Else{
                $D_{a_{\lambda},\lambda}\leftarrow D_{a_{\lambda-1},\lambda-1}$\;
            }
        }
        $u_{\phi}\leftarrow D_{a_{n-1},n-1}$\;
    }
    \Else(~~~~~~~~~~~~~~~~~~~~~//frozen bits){
        $t\leftarrow u_{\phi}$\;
        \For{$\lambda=(n-1):-1:1$}{
            \If{$b_{\lambda}=0$}{
                $t\leftarrow t\oplus D_{a_{\lambda},\lambda}$\;
            }
            \Else{
                $D_{a_{\lambda},\lambda}\leftarrow t$\;
            }
        }
        \If{$b_0=0$}{
            $x_{\phi}\leftarrow x_{\phi}\oplus t$\;
        }
        \Else{
            $x_{\phi}\leftarrow t$\;
            $D_{0,0}\leftarrow t$\;
        }
    }
    \For{$\lambda=1:(n-1)$}{
        \If{$a_{\lambda}=0$}{
            $a_{\lambda}\leftarrow 2^{\lambda}$\;
        }
        $a_{\lambda}--$\;
    }
}
\normalsize
\end{algorithm}

\subsection{Simplification for SPC encoder}
One might think that we need extra bit memory for $b_{\lambda}$ and $a_{\lambda}$ and extra computation for the updating of $a_{\lambda}$. In fact, $\phi$, $b_{\lambda}$ and $a_{\lambda}$ can share the same memory and only updating $\phi$ is enough for all updating. We will show this in the following.

\begin{figure}[!htb]
\centering{\includegraphics[width=50mm]{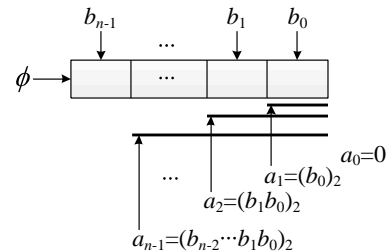}}
\caption{The bits memory for $\phi$, $b_{\lambda}$ and $b_{\lambda}$}
\label{BinaryExpression}
\end{figure}

In hardware implementation, $\phi$ is expressed in binary as $(b_{n-1}\cdots b_1 b_0)_2$, which is shown in Fig.\ref{BinaryExpression}. When $\phi$ is updated, the operations in current horizontal connection are decided by the values of $b_{n-1},\cdots,b_1$ and $b_0$. $(b_{n-1}\cdots b_1b_0)$ will be bitwise visited as one switch to select the corresponding operation for the propagation from right to left (or vice versa).

In layer $\lambda$, $2^{\lambda}$ bits memory are required, which means $a_{\lambda}$ must be a number of $\lambda$ bits. Note that $(b_{\lambda-1}\cdots b_0)_2$ has the same value (in decimal) as $a_{\lambda}$ ($\lambda>0$). Thus, $a_{\lambda}$ can be obtained by selecting $b_{\lambda-1}\cdots b_0$ from $\phi$ without extra memory as shown in Fig.\ref{BinaryExpression}. In layer 0, $a_0$ is fixed to be 0 due to that only one bit is required.

In \textbf{Algorithm \ref{alg:MyEncodingSPC}}, if a full word $\textbf{u}$ is not required, line 15 can be deleted.

\section{Discussion on parallel encoding}
The encoding algorithm described in \textbf{Algorithm \ref{alg:MyEncodingSPC}} works in a serial fashion. To further improve the encoding throughput, we discuss the implementation of a parallel encoding with 2 bits at a time in this section.

\begin{figure}[h]
\centering
\subfigure[]{\label{parallela}\includegraphics[width=0.2\columnwidth]{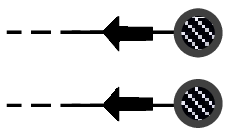}}\hspace{0.03\columnwidth}
\subfigure[]{\label{parallelb}\includegraphics[width=0.2\columnwidth]{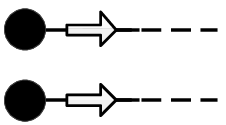}}\hspace{0.03\columnwidth}
\subfigure[]{\label{parallelc}\includegraphics[width=0.2\columnwidth]{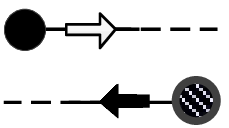}}\hspace{0.03\columnwidth}
\subfigure[]{\label{paralleld}\includegraphics[width=0.2\columnwidth]{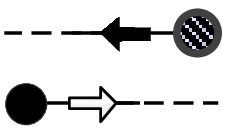}}\hspace{0.03\columnwidth}
\caption{The four cases with top being the $(2\psi)$-{th} bit and bottom being the $(2\psi+1)$-{th} bit: (a) both are user bits, (b) both are frozen bits, (c) frozen bit on top and user bit at the bottom, (d) user bit on top and frozen bit at the bottom.}
\label{parallel}
\end{figure}

Similar to the previous algorithm, the encoding is processed from bottom to top as shown in Fig.\ref{EncodingStructure}. Let $2\psi$ and $2\psi+1$ denote the indices of the two horizontal connections being processed at a time, respectively. Here $\psi\in\{0,1,\cdots,\frac{N}{2}-1\}$. From Fig.\ref{EncodingStructure}, it can be found that there are four different cases of information propagation for two known bits in the encoding process as depicted in Fig.\ref{parallel}. However, it turns out that case (d) never happens for polar codes constructed on symmetric binary memoryless channels (S-BMC).
\begin{proposition}
For $N=2^n~(n\geq 1)$ polar codes with coding structure similar to Fig.\ref{EncodingStructure}, the $(2\psi)$-{th} bit channel, $W_{2\psi}$, must be frozen if the $(2\psi+1)$-{th} bit channel, $W_{2\psi+1}$, is frozen on S-BMC.
\end{proposition}

\begin{IEEEproof}
The result can be obtained by invoking Lemma 5 in [13].
\end{IEEEproof}

We now elaborate on the parallel encoding algorithm and the corresponding memory requirements, where $\phi=2\psi$ and $\psi$ decreases from $(\frac{N}{2}-1)$ to 0. $a_\lambda~(\lambda>0)$ and $b_\lambda$ are obtained in the same way as in Subsection III.B and $a_0$ is updated to be the same as $a_1$. In the following, we use $\textbf{D}_{a_\lambda,\lambda}^{+1}$, $\textbf{u}_{\phi}^{+1}$ and $\textbf{x}_{\phi}^{+1}$ to denote the two-bit vectors $(D_{a_\lambda,\lambda}, D_{a_\lambda+1,\lambda})$, $(u_{\phi}, u_{\phi+1})$ and $(x_{\phi}, x_{\phi+1})$, respectively.

As shown in Fig.\ref{EncodingStructure}, in layer 0 of case (a), with parallel processing, the direct copying operation and XOR operation are performed simultaneously. As such, one additional memory bit, denoted as $D_{1,0}$, is required and then the above operations are done as $\textbf{D}_{0,0}^{+1}\leftarrow(x_{\phi}\oplus x_{\phi+1}, x_{\phi+1})$. In layer $\lambda(>0)$, it can be noticed that the operation types of the $\phi$-th and $(\phi+1)$-th user bits are the same and will be decided according to $b_\lambda$, where the operations of $\textbf{D}_{a_\lambda,\lambda}^{+1}\leftarrow\textbf{D}_{a_{\lambda-1},\lambda-1}^{+1}$ are implemented when $b_\lambda=1$ while the operations of $\textbf{D}_{a_\lambda,\lambda}^{+1}\leftarrow\textbf{D}_{a_\lambda,\lambda}^{+1}\oplus \textbf{D}_{a_{\lambda-1},\lambda-1}^{+1}$ for $b_\lambda=0$. And the operations of $\textbf{u}_\phi^{+1}\leftarrow\textbf{D}_{a_{n-1},n-1}^{+1}$  will be done at the last of case (a).

Note that $\textbf{D}_{0,0}^{+1}$ are idle when the process is done from layer $n-1$ to 1 and we use them as temporary variables in case (b). Then, the process of case (b) is the same as lines 17-22 in \textbf{Algorithm \ref{alg:MyEncodingSPC}} but 2-bits per computation cycle. At the last, $\textbf{u}_{\phi}^{+1}$ is updated with $(D_{0,0}\oplus D_{1,0}, D_{1,0})$ where the new values of $\textbf{D}_{0,0}^{+1}$ after layer 1 are also expected values in the encoding.

Case (c) has two opposite information propagations and still works in the serial mode as \textbf{Algorithm \ref{alg:MyEncodingSPC}} in which $D_{0,0}$ and $D_{1,0}$ are used to replace $t$ in the frozen bit processing part and $D_{0,0}$ in the user bit processing part, respectively.

Consider polar codes with $N=1024$ and code rate 1/2 constructed at signal-to-noise ratio 2dB under additive white Gaussian noisy channel as an illustration, the number of case (a), (b) and (c) are 135, 135 and 242 respectively, that is, 754 horizontal propagations will be implemented with our proposed parallel encoding, which can obtain about 36\% gain in throughput with comparison to \textbf{Algorithm \ref{alg:MyEncodingSPC}}.

From the above description, it can be noted that a new bit memory, i.e. $D_{1,0}$, is introduced in the parallel 2-bit encoding algorithm while the temporary variable $t$ in \textbf{Algorithm \ref{alg:MyEncodingSPC}} is no longer required. Hence, the requirement of bit memory for the parallel 2-bit encoding algorithm is the same as \textbf{Algorithm \ref{alg:MyEncodingSPC}}. Also, it can be easily verified that the number of XOR operations remains unchanged.

Limited to the two opposite information propagations of frozen bits and user bits in the SPC encoding, parallel multiple-bit encoding for SPC is more complex and the storage memory and computation will also increase, which is beyond the discussion of this paper and will be left for future work.

\section{Conclusions}
We have presented an efficient encoding algorithm for SPC which need $N$ bits memory and $\frac{N}{2}\log_2N$ XOR operation, a minimum requirement for the available encoding methods. By sharing memory, our analysis shows the algorithm can be further simplified. To improve encoding throughput, a parallel 2-bit encoding algorithm has also been discussed, which shows the parallel encoding algorithm can be achieved with the same memory bits and XOR operations.


%

\ifCLASSOPTIONcaptionsoff
  \newpage
\fi

\end{document}